\documentclass[prl,aps,twocolumn,superscriptaddress,showpacs,amsmath,amssymb] %tightenlines, twocolumn
{revtex4} \usepackage{dcolumn} \usepackage{bm} \usepackage{graphics}
\usepackage{epsfig}

\def\Hil{\mathcal{H}}
\def\ket#1{|#1\rangle }

\begin{document}

\title{Entanglement in an expanding spacetime}

\author{Jonathan L. Ball}%\email{jonathan.ball@keble.ox.ac.uk}
\affiliation{Centre for Quantum Computation, University of Oxford,
OX1 3PU, U.K.}
\author{Ivette Fuentes-Schuller (n\'ee Fuentes-Guridi)}
%\email{i.fuentesguridi1@physics.ox.ac.uk}
\affiliation{Centre for Quantum Computation, University of Oxford,
OX1 3PU, U.K.} \affiliation{Perimeter Institute for Theoretical
Physics, 31 Caroline Street N, Waterloo N2L 2Y5, Canada.}
\author{Frederic P. Schuller}%\email{fschuller@perimeterinstitute.ca}
\affiliation{Perimeter Institute for Theoretical Physics, 31
Caroline Street N, Waterloo N2L 2Y5, Canada.}
\date{\today}

\begin{abstract}
We show that a dynamical spacetime generates entanglement between
modes of a quantum field. Conversely, the entanglement encodes information
concerning the underlying spacetime structure, which hints at the
prospect of applications of this observation to cosmology. Here we
illustrate this point by way of an analytically exactly soluble
example, that of a scalar quantum field on a two-dimensional
asymptotically flat Robertson-Walker expanding spacetime. We
explicitly calculate the entanglement in the far future, for a
quantum field residing in the vacuum state in the distant past. In
this toy universe, it is possible to fully reconstruct the parameters of the cosmic
history from the entanglement entropy.
\end{abstract}

 % Entanglement production, quantum mechanics, quantum information

\maketitle

The phenomenon of entanglement has attracted much attention in
recent years. Its central importance in the exciting discipline of
quantum information science is undisputable: it has emerged as a
fundamental resource in quantum communication \cite{quantcomm},
quantum cryptography \cite{quantcrypto}, quantum teleportation
\cite{quanttele} and quantum computation \cite{quantcomp}. Recent
effort has begun to translate some of the aforementioned concepts to
the special relativistic setting
\cite{special1,special2,special3,special4,special5,special6} and
recently progress has been made in examining teleportation
\cite{teleport} and entanglement between modes of a quantum field
\cite{ivette} when one of the observers is uniformly accelerated.

However, quantum information remains a discipline that commonly
avoids the conceptual challenges implied by one of the most
fundamental insights of modern physics, namely that spacetime is
dynamical and curved. When first attempting to understand the basic
principles of quantum information, the simplifying assumption of a
flat, or even non-relativistic, spacetime is justifiable. However,
since our ultimate goal is to properly understand the nature of the
universe on a more complete level, and not merely a restricted,
simplified aspect of it, we must necessarily extend our analysis to
the more general curved spacetime domain in which we live. Moreover,
while for many physical systems of interest to quantum information
theory a non-relativistic approximation is sufficient, the latter
entirely fails for massless particles, such as photons, or in the
presence of strong gravitational fields. The advent of precision
measurements in cosmology \cite{cosmo}, for instance, produces a
host of data, whose interpretation will be further enriched by
translation of concepts from quantum information theory to curved
spaces.

In this brief note we consider the effect that an expanding universe has
on the entanglement shared between scalar particles residing in that
spacetime, demonstrating that such a dynamic background structure
actually creates entanglement. Furthermore, we explicitly
demonstrate the fascinating possibility to deduce cosmological
parameters of the underlying spacetime from the entanglement shared
between certain modes of a quantum field.

The observation that entanglement is affected by the underlying
spacetime structure may appear rather startling to those unfamiliar
with quantum field theory on curved spacetime, but is an immediate
consequence of the latter \cite{Birrelldavies}. Indeed, many of the
constructions of quantum theory in flat spacetime, such as the
notion of a particle, only possess limited validity in the general
setting. Despite such conceptual challenges, we show that the
utility of entanglement can be fruitfully extended beyond its usual
domain of non-relativistic quantum information.

We present the required theory in fair generality, but for
simplicity discuss as an explicit example a toy universe in two
spacetime dimensions. More realistic examples differ not in
principle, but only in their analytical complexity.

On a curved $d$-dimensional spacetime $M$ with metric $g$, we
consider a complex valued scalar field $\phi$, whose dynamics is
given by variation of the action
\begin{equation}
S = \int_M d^dx
\sqrt{g}[g^{\mu\nu}(x)\partial_\mu\phi\,\partial_\nu\phi-m^2
\phi^2],
\end{equation}
where $g = (-1)^{d+1} \det g_{\mu\nu}$ and $m$ is a real
positive parameter. The corresponding equation of motion is
\begin{equation}
 (\Box + m^2)\phi(x)=0, \label{eq:fecurved}
\end{equation}
where
$\Box\phi=\partial_\mu(\sqrt{g}g^{\mu\nu}\partial_\nu\phi)/\sqrt{g}$.
Due to the linearity in $\phi$, the space of solutions constitutes a
vector space, which is made into a Hilbert space $\Hil$ by equipping
it with the time-independent inner product
\begin{equation}
\label{eq:product}
(\phi,\psi) = i \int_\Sigma
d\Sigma^\mu [\psi^* \partial_\mu \phi - \phi \partial_\mu
\psi^*],
\end{equation}
where $\Sigma$ is a spacelike Cauchy surface. As in any complex
Hilbert space, we can find an orthonormal basis of the form
$\{u_p,u_p^*\}$ such that
\begin{equation}
(u_p,u_q) = -(u_p^*, u_q^*) = \delta_{pq} \textrm{ and }
(u_p,u_q^*)=0,
\end{equation}
where $p, q$ are in some (possibly continuous) index set $\mathcal{P}$. 
In general, no particular splitting of the basis into $u_p$ and
$u_p^*$ is distinguished. In a spacetime with time translation
symmetry, however, we have a timelike Killing vector field $K$,
satisfying $\mathcal{L}_K g = 0$, where $\mathcal{L}$ denotes the
Lie derivative. The corresponding energy conservation allows one to
meaningfully classify solutions of (\ref{eq:fecurved}) into positive
and negative frequency solutions, if they are eigenfunctions of $i
\mathcal{L}_K$ with positive or negative eigenvalues, respectively.

Canonical quantization of the theory consists of promoting the field
$\phi$ to an operator field
\begin{equation}
\hat\phi(x) = \int_\mathcal{P} [a_p^- u(p) + a^+_p u^*(p)],
\end{equation}
where $a^-_p$ and $ a^+_p$ act on the bosonic Fock space
\begin{equation}
 \mathcal{F} = \mathbb{C} \oplus \Hil \oplus
(\Hil\circ\Hil) \oplus (\Hil\circ\Hil\circ\Hil) \dots
\end{equation}
by linear extension of their action on $\Hil^{\circ n}$,
\begin{equation}
\begin{split}
a^-_p \ket{q_1}....\ket{q_n} &= \sum_{i=1}^n \ket{q_1}....\langle
p | q_i \rangle....\ket{q_n} \in \Hil^{\circ (n-1)},\\
a_p^+ \ket{q_1}....\ket{q_n} &= \ket{p}\ket{q_1}....\ket{q_n}
\qquad\qquad \in \Hil^{\circ (n+1)},\\
\end{split}
\end{equation}
where the order of the kets is irrelevant due to the symmetric
tensor product $\circ$. The Fock space $\mathcal{F}$ inherits the
inner product (\ref{eq:product}) from $\Hil$, from which it
immediately follows that $ a^+_p = ( a^-_p)^\dagger$, where
$\dagger$ denotes the adjoint with respect to the inner product on
$\mathcal{F}$. We will from now on denote $ a^-_p$ simply as $ a_p$.
From the above construction it follows that
\begin{equation}
   [a_p^\dagger, a_q] = \delta_{pq} \textrm{ and } [ a_p, a_q] = [
a_p^\dagger, a_q^\dagger] = 0.
\end{equation}
These commutation relations are often stated as the essence of the
canonical quantization procedure for bosons. If the $u_p$ are
positive frequency solutions with respect to some timelike Killing
vector field $K$, the operators $a_p^\dagger$ and $a_p$ may be
meaningfully interpreted as creation and annihilation operators for
a particle excitation (of mass $m$, momentum $p$, and energy
$\sqrt{p^2 + m^2}$), if the vacuum state $\ket{0}\equiv 1_\mathbb{C}
\oplus 0_\mathcal{H} \oplus \dots \in \mathcal{F}$ is interpreted as
the no-particle state. Note that from this definition it immediately
follows that $ a_p\ket{0} = 0$ for all $p\in\mathcal{P}$.

On a generic curved spacetime, there exists no global timelike
Killing vector field $K$, so that in general no meaningful particle
interpretation can be attached to a state of the quantum field.
However, if in some region of the spacetime under consideration,
there exists a local Killing vector field, we will exploit the fact
that a particle interpretation exists in that region. In the absence
of a global Killing vector field, the same quantum state then
generically has particle interpretations varying between those
regions that feature their own individual local Killing vector
fields, and thus different positive and negative frequency modes.
The corresponding change from one set of positive and negative
frequency modes $\{u_p,u_p^*\}$ to another set $\{\bar u_p, \bar
u_p^*\}$ is simply a change of the Hilbert space basis. As creation
and annihilation operators are defined with respect to a specific
mode decomposition, changing the latter induces the Bogolubov
transformations
\begin{equation}
   \bar  a_p = \int_{q\in\mathcal{P}}
[\alpha_{p q}^*  a_q - \beta_{p q}^*  a_q^\dagger],
\end{equation}
parameterized by the projection coefficients $\alpha_{p q} = (\bar
u_p, u_q)$ and $\beta_{p q} = -(\bar u_p, u_q^*)$ of the basis
change in $\Hil$. In the case of non-vanishing $\beta_{pq}$, it
follows that the vacua $\ket{0}$ and $\ket{\bar 0}$, defined with
respect to the different mode decompositions, are inequivalent. As a
consequence, the particle concept, so familiar and widely used in
discussions of conventional quantum information, is a more intricate
one on curved spacetime.

The majority of explicit calculations in quantum field theory on
curved spacetime are notoriously difficult. However, there exist
certain specific models of the spacetime structure that are exactly
analytically soluble. We shall consider one such simple model that
is asymptotically flat in the distant past and far future, as this
allows us to easily illustrate the main principles and consequences
of examining entanglement in curved spacetime without having to
resort to approximations or numerical solutions.

Specifically, we consider a two-dimensional Robertson-Walker
expanding spacetime with line element
\begin{equation}
ds^2=C(\tau)(d\tau^2-dx^2),
\end{equation}
where $\tau$ is the conformal time and the conformal scale factor is
given as
\begin{equation}
 C(\tau) = 1 + \epsilon(1 + \tanh \sigma\tau),
\end{equation}
with positive real parameters $\epsilon$ and $\sigma$, controlling
the total volume and rapidity of the expansion. This describes a toy
universe undergoing a period of smooth expansion. In the distant
past and far future, the spacetime becomes Minkowskian since
$C(\tau)$ tends to $1+2\epsilon$ and $1$ as
$\tau\rightarrow\pm\infty$, respectively. As a consequence, the
vector field $K=\partial/\partial\tau$ has the Killing property in
both the asymptotic in-region ($\tau\rightarrow-\infty$) and
out-region ($\tau\rightarrow\infty$), but not for finite $\tau$. In
the asymptotic regions it is possible to sensibly discuss the
particle content of a scalar field; in the intermediate region,
however, the concept of a particle breaks down.

In order to find the solutions to the Klein-Gordon equation
Eq.~(\ref{eq:fecurved}) on this spacetime, we note that $C(\tau)$ is
independent of $x$. We exploit the resulting spatial translational
invariance and separate the solutions into
\begin{equation}
 \phi_p(\tau,x)=(2\pi)^{-1/2}e^{i px}\chi_p(\tau),
\end{equation}
so that $\mathcal{P}=\mathbb{R}$ in this example. Inserting this
into the Klein-Gordon equation, we obtain a simple differential
equation for $\chi_p(\tau)$ which may be solved in terms of
hypergeometric functions \cite{Birrelldavies}. We then apply the Lie
derivative $\mathcal{L}_K$ to the solutions $\phi_p$, in order to
identify the normalized modes $\bar u_p$ which behave like positive
frequency modes in the remote past, and the positive frequency modes
$u_p$ in the far future, respectively:
\begin{eqnarray}
\bar u_p(\tau,x)&\stackrel{\longrightarrow}{{}_{\tau\rightarrow
-\infty}}&
(4\pi\omega_\text{in})^{-1/2}e^{i(px-\omega_\text{in}\tau)},\nonumber\\
u_p(\tau,x)&\stackrel{\longrightarrow}{{}_{\tau\rightarrow
+\infty}}&
(4\pi\omega_\text{out})^{-1/2}e^{i(px-\omega_\text{out}\tau)},
\end{eqnarray}
where the angular frequencies have the form
\begin{eqnarray}
\omega_\text{in}&=&[p^2+m^2]^{1/2},\nonumber\\
\omega_\text{out}&=&[p^2+m^2(1+2\epsilon)]^{1/2},\nonumber\\
\omega_\pm&=&\frac{1}{2}(\omega_\text{out}\pm\omega_\text{in}).
\end{eqnarray}
We shall henceforth denote all quantities related to the in-region
using a bar, and those referring to the out-region without a bar. A
consequence of the linear transformation properties of
hypergeometric functions is that the Bogolubov transformations
associated with the transformation from $\{\bar u_p, \bar u_p^*\}$
to $\{u_p, u_p^*\}$ take the simple form
\begin{equation}
\bar{a}_s=\alpha^*_s a_s - \beta^*_s a^\dagger_{-s},
\label{Eq:bogsimp}
\end{equation}
so that mixing occurs only between states labelled by $s$ and
$-s$.

Now consider the case where the ingoing scalar field is in a vacuum
state $\prod_{s\in\mathbb{R}{}^{+}}
|\bar{0}\rangle_s|\bar{0}\rangle_{-s}$, with no excitations present
in any of the modes (from the point of view of an inertial observer
in the in-region). Because of the simple mixing properties of the
Bogolubov transformations, we may focus solely on a component
$|\bar{0}\rangle_k|\bar{0}\rangle_{-k}$ of the input state (disregarding all other modes, by tracing the total density matrix over them, yields an overall factor of unity), and
express this component in terms of the modes in the out-region. As a
pure state of a bi-partite system, this can be written as a Schmidt
decomposition
\begin{equation}
|\bar{0}\rangle_k|\bar{0}\rangle_{-k}=\sum^\infty_{n=0}c_n|n\rangle_k|n\rangle_
{-k}, \label{Eq:intoout}
\end{equation}
where $n$ labels the number of excitations in the field mode $k$ (as
seen by an inertial observer in the out-region) and the coefficients
$c_n$ are real. An explicit expression for the Schmidt coefficients
$c_n$ can be obtained by applying  Eq.~(\ref{Eq:bogsimp}) to
Eq.~(\ref{Eq:intoout}):
\begin{equation}
0=\bar{a}_k|\bar{0}\rangle_k|\bar{0}\rangle_{-k}=\left(\alpha^*_ka_k-
\beta^*_ka^\dagger_{-k} \right)\sum_{n=0}^\infty
c_n|n\rangle_k|n\rangle_{-k}.
\end{equation}
A simple relabelling of the mode excitation number then allows
one to deduce that
\begin{equation}
c_n=\left(\frac{\beta^*_k}{\alpha^*_k}\right)^n c_0,
\end{equation}
whilst taking the inner product of Eq.~(\ref{Eq:intoout}) with its
hermitian conjugate yields the following value for the first Schmidt
coefficient:
\begin{equation}
c_0=\sqrt{1-\left|\frac{\beta_k}{\alpha_k}\right|^2}.
\end{equation}
Thus Eq.~(\ref{Eq:intoout}) shows that a state which is interpreted
as a vacuum in the in-region appears as a state with particle
excitations in the out-region. We must interpret this fact as the
creation of particles as a direct result of the cosmic expansion.
Recall however that in the interim region, when our toy universe is
undergoing expansion, no sensible notion of a particle exists.

We are now in a position to apply familiar methods of quantum
information theory to extract information about the entanglement of
the state Eq.~(\ref{Eq:intoout}). It is a simple matter to construct
the asymptotic output state bipartite density matrix $\varrho =
|\bar 0\rangle_{-k}|\bar 0\rangle_{k}\,{}_{k}\langle\bar
0|{}_{-k}\langle\bar 0|$, describing excitations in the scalar field
modes $k$ and $-k$. Because the Schmidt coefficients in Eq.
(\ref{Eq:intoout}) are non-zero, the in-vacuum is entangled from the
point of view of an observer in the out-region. Since the density
matrix $\varrho$ describes a pure state, the von Neumann entropy $S$
of the reduced density matrix
\begin{equation}
\varrho_k = \sum_{m=0}^{\infty} {}_{-k}\langle m | \varrho
|m\rangle_{-k}
\end{equation}
presents a well-defined measure for this entanglement
of the modes $k$ with the modes $-k$. One finds, after some algebra,
the entanglement
\begin{equation}
\label{Eq:entang}
S = - \textrm{Tr}(\varrho_k \log_2
\varrho_k) =  \log_2 \frac{\gamma^{\gamma/(\gamma-1)}}{1-\gamma},
\end{equation}
where
\begin{equation}
\label{gammaS}
\gamma=\left|\frac{\beta_k}{\alpha_k}\right|^2=\frac{\sinh^2(\pi\omega_-/
\sigma)}{\sinh^2(\pi\omega_+/\sigma)}
\end{equation}
depends on the cosmological parameters $\epsilon$ and $\sigma$, and
the momentum $k$ of the selected modes. This means that the
expansion of the universe creates entanglement between massive modes
of opposite momenta. Modes of a massless quantum field do not get
entangled, due to the conformal flatness of the model studied here.
Although this total decoupling of massless modes is an artefact of
the specific example studied here, it illustrates that the massive
and massless case are generically qualitatively different, and
non-relativistic intuition fails entirely. Since $0\leq\gamma<1$,
the entanglement is monotonically increasing in $\gamma$, and we may
invert Eq.~(\ref{Eq:entang}) to obtain $\gamma(S)$.

Indeed, the entanglement between the field modes encodes the entire
information about the underlying spacetime. For light particles, a direct relation between the cosmological parameters and the degree of entanglement can be obtained. To see this explicitly,
assume that a universe of the type discussed above, with
cosmological parameters $\sigma$ and $\epsilon$,  is populated by a
particle species of mass $m \ll 2\sigma\epsilon^{-1/2}$. Then we can
consider quanta of energy $E_p=\sqrt{p^2+m^2}$ such that $m
\sqrt{\epsilon} \ll E_p \ll 2 \sigma$, and we obtain the frequencies
\begin{eqnarray}
\omega_+ &\approx& E_p \ll \sigma,\\
\omega_- &\approx& \omega_+ \frac{m^2}{2 E_p^2} \epsilon \ll \sigma,
\end{eqnarray}
such that the cosmological expansion parameter $\epsilon$ can be
determined, by expanding Eq.~(\ref{gammaS}) to leading order in
$\epsilon$, as a monotonically increasing function of the
entanglement $S$:
\begin{equation}
\epsilon \approx \frac{2 E_p^2}{m^2} \sqrt{\gamma(S)}.
\end{equation}
The cosmological parameter $\sigma$ can be determined from the respective entanglement of 
two modes of slightly different energy. More precisely, one finds by differentiating Eq. (\ref{gammaS}) with respect to the particle energy $E$ that
\begin{equation}
  \sigma \approx \frac{\pi}{2} \left(\frac{1+\gamma(S)}{-\frac{E}{4}\frac{d}{dE}\ln \gamma(S) - 1}\right)^\frac{1}{2} E,
\end{equation}
using the above approximations. Thus, in our simple toy universe, the
entanglement of massive states carries the complete information
about the cosmological parameters.

In a more realistic four-dimensional setting, neutrinos present a
natural candidate for such very light \cite{neutrinomass} and
approximately only gravitationally interacting particles. The
required Bogolubov transformations for a four-dimensional
Friedmann-Robertson-Walker universe have been calculated
\cite{stefan}, and for spacetime manifolds admitting a spin
structure, the Dirac equation for spin-$1/2$ fermions can be written
down and quantized. Although this more realistic setting is
analytically considerably more involved, the qualitative features
are the same. In particular, DeWitt \cite{dewitt} gives the general
vacuum to multi-particle production and annihilation amplitudes
needed in order to generalize our simple example. The degree of
entanglement of neutrinos therefore likely carries information
concerning our cosmic history, a fascinating speculation.

Our analysis assumed that the quantum field is in a vacuum state
from the point of view of an inertial observer in the distant past.
While without further assumptions this presents a natural departure
point, one could instead calculate the fate of a non-vacuum state,
or even an initially entangled state. The formalism employed above
is directly applicable also to those cases. We have neglected,
however, any back reaction of the quantum field on the spacetime
through the Einstein equations. If the dynamics of the universe are
not significantly driven by the quantum field under consideration,
this provides a reasonable approximation.

In conclusion, by extending our understanding of entanglement to a
curved spacetime background, we have learned that the latter has a
significant effect. We showed that an expanding spacetime generates
entanglement between certain modes of an only gravitationally
interacting scalar field. While this is not an unexpected result
within quantum field theory on curved spacetime, we made the
interesting observation that, conversely, information on the
underlying spacetime can be recovered from the entanglement of very
light particles. We exemplified this by pointing out how all
cosmological parameters of a toy expanding universe can be extracted
from quantum correlations. We will show elsewhere in detail how our
treatment can be applied to other spacetimes that are either more
realistic or otherwise of interest. Far from being of mere interest
to quantum information theory, a further exploration of entanglement
in a curved spacetime therefore promises to contribute to a
cross-fertilization between gravity and quantum information. Recent
progress on generally covariant quantum field theory supplies
valuable new tools for such investigations \cite{verch}.

\begin{acknowledgments}
The authors would like to thank Stefan Hofmann for useful
discussions. JLB acknowledges financial support from Keble College
in Oxford and EPSRC, and thanks Perimeter Institute for its generous
hospitality.
\end{acknowledgments}


\begin{thebibliography}{99}
\bibitem{quantcomm} C.H. Bennett, P.W. Shor, J.A. Smolin and A.V.
Thapliyal, Phys. Rev. Lett. {\bf 83}, 3081-3084 (1999); C.H. Bennett
and S.J. Wiesner, Phys. Rev. Lett. {\bf 69}, 2881 (1992).

\bibitem{quantcrypto} A.K. Ekert, Phys. Rev. Lett. {\bf 67}, 661-663
(1991).

\bibitem{quanttele} C.H. Bennett, G. Brassard, C. Cr\'epeau, R. Josza, A. Peres,
W.K. Wooters, Phys. Rev. Lett. {\bf 70}, 1895 (1993).

\bibitem{quantcomp} A.M. Steane, Rep. Prog. Phys. {\bf 61}, 117-173
(1998).

% \bibitem{ekertbook} \lq\lq The Physics of Quantum Information'',
% Springer-Verlag 2000, D. Bouwmeester, A. Ekert, A. Zeilinger
% (Eds.).

\bibitem{special1} D.R. Terno and A. Peres, Rev. Mod. Phys. {\bf 76}, 93
(2004).

\bibitem{special2} A. Peres, P.F. Scudo and D.R. Terno,
Phys. Rev. Lett. {\bf 88}, 230402 (2002).

\bibitem{special3} P.M. Alsing and G.J.
Milburn, Quant. Inf. Comp. {\bf 2}, 487 (2002).

\bibitem{special4} R.M. Gingrich and
C. Adami, Phys. Rev. Lett. {\bf 89}, 270402 (2002).

\bibitem{special5} J. Pachos and E. Solano, \textit{QIC Vol. 3, No. 2, pp.115}
(2003).

\bibitem{special6} P. Calabrese and J. Cardy, JSTAT 0406 (2004)
P002.

\bibitem{teleport}P.M. Alsing and G.J. Milburn, Phys. Rev. Lett.
{\bf 91}, 180404 (2003).

\bibitem{ivette} I. Fuentes-Schuller and R.B. Mann,
quant-ph/0410172.

\bibitem{cosmo} D.N. Spergel et al., Astrophys. J. Suppl. {\bf 148} (2003)
175.

% \bibitem{cosmo2} A.G. Riess et al., Astron. J. {\bf 116} (1998) 1009

\bibitem{Birrelldavies}
N.D. Birrell and P.C.W. Davies, Quantum fields in curved space, CUP
1994.

\bibitem{neutrinomass} Y. Fukuda et al., Phys. Rev. Lett. {\bf 81} 1562
(1998).

\bibitem{stefan} D. Polarski and A. Starobinski, Class. Quantum Grav. {\bf
13}, 377 (1996).

\bibitem{dewitt} B.S. DeWitt, Phys. Reports {\bf 19} 295 (1975).

\bibitem{verch} R. Verch and R. F. Werner, to appear in Rev. Math. Phys.

\end{thebibliography}
\end{document}